\documentclass[aps,prd,nofootinbib,twocolumn,preprintnumbers,superscriptaddress]{revtex4-1}
\pdfoutput=1

\usepackage[dvipsnames]{xcolor}
\usepackage{graphicx}
\usepackage{dcolumn}
\usepackage{bm}
\usepackage{hyperref}
\usepackage{multirow}
\usepackage{color}
\usepackage{comment}
\usepackage{amsmath,amsfonts,bbm,mathrsfs,amssymb}
\usepackage{slashed,cancel}
\usepackage{multirow}
\usepackage[normalem]{ulem} 
\usepackage{ytableau}
\usepackage{mathtools}
\usepackage{xcolor}

\setcitestyle{square}

\definecolor{myblue}{rgb}{0.152941176,0.549019608,0.670588235}
\definecolor{newred}{cmyk}{0,1,1,0.2}
\definecolor{newblue}{cmyk}{1,1,0,0.1}
\hypersetup{
    pdftoolbar=true,        
    pdfmenubar=true,        
    pdffitwindow=false,     
    pdfstartview={FitH},    
    pdftitle={Production of dark sector particles via resonant positron annihilation on atomic electrons},
    pdfauthor={F. Arias Aragon, L. Darme, G. Grilli di Cortona, E. Nardi},
    pdfkeywords={X17} {dark photon} {atomic electron} {crystal} {resonant},
    pdfnewwindow=true,
    colorlinks=true,
    linkcolor=myblue,
    citecolor=myblue,
    filecolor=myblue,
    urlcolor=myblue,
    linktocpage=true
}

\def\equationautorefname~#1\null{Eq.\,(#1)\null}

\def\eq#1{{Eq.~(\ref{#1})}}

\newcommand{\be}{\begin{equation}}
\newcommand{\ee}{\end{equation}}
\newcommand{\ba} {\begin{equation}\begin{aligned}}
\newcommand{\ea} {\end{aligned}\end{equation}}
\newcommand{\bg} {\begin{equation}\begin{gathered}}
\newcommand{\eg} {\end{gathered}\end{equation}}

\newcommand{\sL}{\mathscr{L}}

\newcommand{\cM}{\mathcal{M}}

\newcommand{\bea}{\begin{eqnarray}}
\newcommand{\eea}{\end{eqnarray}}

\begin{document}

\title{Production of dark sector particles via resonant positron annihilation on atomic electrons}
\author{Fernando~Arias-Arag\'on}
\email{fernando.ariasaragon@lnf.infn.it}
\affiliation{Istituto Nazionale di Fisica Nucleare, Laboratori Nazionali di Frascati, Frascati, 00044, Italy}

\author{Luc Darmé}
\email{l.darme@ip2i.in2p3.fr}
\affiliation{Institut de Physique des 2 Infinis de Lyon (IP2I), UMR5822, CNRS/IN2P3, F-69622 Villeurbanne Cedex, France}

\author{Giovanni Grilli di Cortona}
\email{giovanni.grilli@lngs.infn.it}
\affiliation{Istituto Nazionale di Fisica Nucleare, Laboratori Nazionali del Gran Sasso, Assergi, 67100, L'Aquila (AQ), Italy}
 
\author{Enrico Nardi}
\email{enrico.nardi@lnf.infn.it}
\affiliation{Istituto Nazionale di Fisica Nucleare, Laboratori Nazionali di Frascati, Frascati, 00044, Italy}
\affiliation{Laboratory of High Energy and Computational Physic, HEPC-NICPB, R\"avala 10, 10143, Tallin, Estonia}
\date{\today}

\begin{abstract}
Resonant positron annihilation on atomic electrons provides a powerful method to search for light new particles coupled to $e^+e^-$. Reliable estimates of production rates require a detailed characterization of electron momentum distributions. We describe a general method that harnesses the target material Compton profile to properly include  electron velocity effects in resonant annihilation cross-sections. We additionally find that high $Z$ atoms can efficiently act as particle physics accelerators, providing a density of relativistic electrons that allows to extend by several times the experimental mass reach.
\end{abstract}

\maketitle


\noindent
\section{Introduction} 
Established phenomena like dark matter, the cosmological baryon asymmetry and neutrino masses, which remain unexplained within the Standard Model (SM), provide compelling evidence for the necessity of new physics. Physics beyond the SM may eventually manifest as an entirely novel sector comprising both new particles and interactions. The new states do not need to be particularly heavy to have so far eluded detection; their masses could well be within experimental reach, provided they couple sufficiently feebly to the SM sector. 

New light particles with feeble couplings to electron and positrons 
can be effectively searched for by harnessing intense positron beams 
impinging on fixed targets. This strategy becomes particularly powerful if the conditions for resonant $e^+e^- $ annihilation into the new states  
can be engineered, since this would yield  a huge enhancement in the production rates. Indeed, the initial proposal for leveraging this 
strategy~\cite{NewDirections,Nardi:2018cxi}  has already  
garnered significant attention within the community~\cite{Marsicano:2018krp,
Celentano:2020vtu,Battaglieri:2021rwp,Andreev:2021fzd,Battaglieri:2022dcy,Darme:2022zfw,Alves:2023ree}. 

Resonant production of new particles requires scanning over suitable center-of-mass energy ranges. This can be achieved in two ways, depending on the characteristics of the target: (i) For thin targets with low nuclear charge, where positron energy losses within the material are negligible, the beam energy must be adjusted incrementally in small steps to continuously span the desired range (see e.g. Ref.~\cite{Darme:2022zfw}); (ii) for thick targets of large nuclear charge, the beam energy can be kept fixed, as the in-matter positron energy losses ensure 
a continuous energy scan~\cite{Nardi:2018cxi}.
This applies also to secondary positrons produced  in electromagnetic showers~\cite{Marsicano:2018krp,Marsicano:2018glj,Celentano:2020vtu,Darme:2022bew} initiated by electron or proton beams.
In all cases, a detailed characterization of the momentum distribution of atomic electrons is mandatory to derive reliable estimates of resonant production rates and signal shapes. However, to date, most analyses rely on the simplifying assumption  of electrons being at rest. One exception is the original paper~\cite{Nardi:2018cxi} where data from the Doppler broadening of the 511\,keV photon line from  annihilation of stopped positrons~\cite{Ghosh:2000} were used to  account for  electron velocity effects in tungsten.
 While this  approach is reasonable, the data considered in~\cite{Nardi:2018cxi} describe more properly  the annihilation probability distribution of positrons {\it at rest} as a function of the electron momentum, rather than directly the  electrons momentum distribution~\footnote{Modeling that goes beyond the free-electron-at-rest approximation has been recently incorporated in the computer package DMG4~\cite{Oberhauser:2024ozf} to simulate new particles production  in fixed target experiments. Upon completion of this work,   
Ref.~\cite{Plestid:2024xzh} appeared where atomic binding corrections in 
lepton-target electron scattering are analyzed. Both these works rely on the virial theorem to estimate electron kinetic energies. However, 
average electron velocities do not capture the large effects discussed in this paper.}.

In this letter we argue that  the Compton profile (CP) (see e.g.~\cite{ComptonProfilesReview} for a review) provides an accurate description of the electron momentum distribution for any given material. 
We describe a prescription for harnessing the CP of a target material to properly incorporate the effects of  electron velocities in the cross-section for resonant annihilation 
We emphasize that, since the problem at hand involves complicated aspects of 
solid-state physics, it is crucial to rely on quantities that are experimentally measured. This ensures that theoretical calculations can be directly validated against data. As a concrete example,~we~consider resonant searches for the elusive $X_{17}$ boson, proposed to explain the anomalies observed in the angular correlation spectra in $^8$Be, $^4$He and $^{12}$C nuclear transitions~\cite{Krasznahorkay:2015iga,Krasznahorkay:2018snd,Krasznahorkay:2021joi}.
We illustrate the importance of electron velocity effects by estimating the sensitivity of the PADME experiment~\cite{Raggi:2014zpa,Raggi:2015gza} for  $X_{17}$ searches, both in the case of a carbon ($Z=6$) thin target ($100\,\mu$m) (for which data have already been taken and the analysis is ongoing) and for a tungsten ($Z=74$) thick target ($5\,$cm). We then compare our results with  previous studies~\cite{Darme:2022zfw,Nardi:2018cxi}.  
We anticipate that for the carbon target the effects of electron velocities mainly translate into a certain loss of sensitivity. This is due to the fact that the signal gets spread over a larger range of energies so that the signal-to-noise ratio is decreased. 

In the case of the  high $Z$ target we  observe instead an impressive extension of the reach in mass, by about a factor of four. This is because of head-on collisions with  high-momentum electrons in the tail of the momentum distribution allow to reach much higher center-of-mass (CoM) energies than in the case of electrons at rest.
This result represents an important finding: it can inspire the conception of suitable  experimental setups that,  by leveraging this phenomenon, 
will significantly enhance the mass reach of  searches for new particles with positron beams.

\noindent
\section{Resonant cross section}
In this work we focus on $2\xrightarrow{}1$ processes where a particle $X$ is resonantly produced via positron annihilation on  atomic electrons. 
The differential cross section for positron annihilation  
off an electron with orbital quantum numbers collectively labeled by $q$ can be written as:
\begin{eqnarray}
\label{eq:sigma1}
d\sigma_q &=&\frac{d^3p}{(2\pi)^3} \, \int\frac{d^3k_A}{(2\pi)^3}\int \frac{d^3k_B}{(2\pi)^3} \,\times\nonumber\\
&&\frac{\left|\phi_{A,q}(\bm{k}_A)\right|^2\left|\mathcal{M}(k_A,k_B\rightarrow p)\right|^2\left|\phi_B(\bm{k}_B)\right|^2}{2E_X2E_{k_A}2E_B\left|v_A-v_B\right|}\times\nonumber\\
&&(2\pi)^4\delta(E_{A}+E_B-E_X)\delta^{(3)}(\bm{k}_A+\bm{k}_B-\bm{p}).
\end{eqnarray}
where the subscripts $A$ and $B$ label respectively electron and positron quantities, 
and $\phi(\bm{k}),\ \bm{k}$ denote  their wave function and momentum.  $E_{k_A}=\sqrt{k_A^2+m_e^2}$ appearing in the denominator 
corresponds to the normalization of free electron states, while $E_A=m_e$ in the $\delta$-function is the energy of the bound electron (neglecting its binding energy). Note that a subscript $q$ on $E_A$, $v_A$ and $\bm{k}_A $  is left understood. 
Finally, $p$ and  $E_X$ denote the   momentum and energy of the final $X$ particle.
We refer to Appendix~\ref{app:CS} for more details. 

Let us now treat the  positrons in the beam as free particles with a well defined momentum $\bm{p}_B$, in which case the 
wave function satisfies: 
\be
\label{eq:conditionfree}
\int \frac{dk_B^3}{(2\pi)^3} \left|\phi_B(\bm{k_B})\right|^2 = 1,\,\, \left|\phi_B(\bm{k_B})\right|^2 = (2\pi)^3 \delta^{(3)}(\bm{p}_B-\bm{k}_B) \, .
\ee
In contrast, atomic electrons are not free but confined in space, which implies 
that a certain probability distribution is associated with their momenta. 
Let us now introduce the electron momentum density function $n(\bm{k}_A) $
normalised to the  atomic number $Z$ of the target atoms: 
\be\label{eq:MDF}
n(\bm{k}_A) =\sum_{q} \left|\phi_{q}(\bm{k}_A)\right|^2,   \,\,   
\int\frac{ d^3k_A}{{(2\pi)^3}} \; n(\bm{k}_A) =  Z \ .
\ee
The  electron momentum density distribution $n(\bm{k}_A) $ can be directly related to the CP, that is measured from the Doppler shift of scattered photons in the Compton process $e^-+\gamma \to e^- + \gamma$. 
This is possible because the time scale for the Compton interaction is much shorter than the time scale needed for the spectator electrons to rearrange in a new configuration, so that the initial and final state electrons feel the same potential.  
Thus, in this {\it impulse} approximation, the effect of the binding energies $u_q <0$  cancels out.  
However, for the process $e^+e^- \to X$ the boundary conditions are different, since the $X$  final state does not feel any Coulomb potential and  $u_q$ plays a role in energy conservation.  
Recalling that the positron beam  has an intrinsic energy spread $\sigma_B = \delta_B E_B$, 
it is easy to see  that shifts in the CoM become relevant only when $u_q \gtrsim m_e \delta_B \simeq 2.6\,$keV, 
where we have assumed a typical energy spread $\delta_B = 0.5\%$. 
Thus for low $Z$ materials, and for the outer shell of high $Z$ materials 
(for tungsten, up to $n\geq 3 $), binding energy effects 
can be neglected. 
Inner shells of high $Z$ materials can have large biding energies (for 
tungsten $u_{1s}\! \sim\! 70\,$keV,  $u_{2s,2p}\! \sim\! 10 -12\,$keV). 
However, their contribution to $n(\bm{k}_A)$ at low/medium $\bm{k}_A$ values is subdominant with respect to the contributions of all other electrons in the outer shells. 
At large values of $\bm{k}_A$ the contribution of the inner shells dominates.  However, in this
region, the momenta of the inner electrons can reach values
$k_A\gtrsim m_e$,   so that the
corrections from $u_q\neq 0$ to the c.m. energy 
remain small.\footnote{A prescription to account for binding energies corrections is given in Appendix~\ref{app:binding}.}
Taking $E_A \simeq m_e$ allows to sum \eq{eq:sigma1} over  $q$ to obtain:
\begin{eqnarray}
\label{eq:sigmasumq}
d\sigma &=&\frac{d^3p}{(2\pi)^3} \int\frac{d^3k_A}{(2\pi)^3}  \frac{n(\bm{k}_A)\left|\mathcal{M}(k_A,p_B\rightarrow p)\right|^2}{8 E_X \left|E_Bk_A^z-E_{k_A}p_B^z\right|}\times\nonumber \\
&& (2\pi)^4\delta(E_{A}+E_B-E_X)\delta^{(3)}(\bm{k}_A+\bm{p}_B-\bm{p}),
\label{eq:cs}
\end{eqnarray}
where we have used $\left|v_A-v_B\right|E_{k_A}E_B~=~\left|E_Bk_A^z-E_{k_A}p_B^z\right|$.
We now introduce polar coordinates referred to the beam axis $z$,  with 
$\theta_A,\phi_A$ denoting the polar and azimutal angles of $\bm{k}_A$. Integrating over $d^3p$, we can eliminate three delta functions, after which the  conservation conditions read:
\be
E_A +E_B = E_X=\sqrt{k_A^2+p_B^2+2k_Ap_B x+m_X^2}\,,
\label{eq:EX}
\ee
with $x=\cos\theta_A$.
By leveraging the remaining delta function, we finally obtain:
\be
\label{eq:dCScrysFinal}
\frac{d^2 \sigma}{d k_A d \phi_A} = \frac{|\mathcal{M}|^2}{32\pi^2}\, \frac{k_A n(k_A,\phi_A,x_0) }{p_B|E_B k_A x_0(k_A)-E_{k_A} p_B|}.
\ee
Energy conservation implies  
\begin{align}
x_0(k_A)& 
= \frac{2E_A E_B+2m_e^2-m_X^2-k_A^2 }{2k_Ap_B},
\end{align}
and the electron momentum $k_A$ lies in the range:
$$
    k_A^\mathrm{max,min} =  \left|p_B \pm \sqrt{ \left(E_A + E_B \right)^2-m_X^2} \right|\,. \nonumber
$$


%
Let us now consider a vector particle of mass $m_X$ that couples 
to electrons through the following interaction:
\be
\label{eq:Lag}
\sL_{X}\subset g_V\, X_\mu \bar{e}\gamma^\mu e,
\ee
where $g_V$ is the coupling constant. This interaction describes also 
dark photon (DP) models with a kinetic mixing parameter $\epsilon = g_{V}/e$. 
The spin-averaged free matrix element for  
the resonant  process ${e^+}~{e^-}~{\rightarrow}~{X}$  can be written 
generically in terms of products of four momenta (denoted by capital letters) 
as follows:
\be
\left|\cM\right|^2=g_V^2\left(3m_e^2+K_A\!\cdot\! P_B+\frac{2}{m_X^2}K_A\!\cdot\! P_X\, P_B\!\cdot\! P_X\right),
\label{eq:Msq0}
\ee
where the four-momenta 
correspond to free one-particle states, with 
three-momenta satisfying the conservation condition 
$\bm{k}_A+\bm{p}_B =\bm{p}$, and with time-like component for the electron momentum 
 $E_{k_A}=\sqrt{\bm{k}_A^2+m^2}$ 
(see Appendix~\ref{app:CS} for details). 
In the approximation of electron at rest, \eq{eq:Msq0}
reduces to 
\be
\label{eq:Matrix}
\left|\mathcal{M}_0\right|^2=g_V^2(m_X^2+2m_e^2)\,,
\ee
which does not depend on $k_A$, and can then be factored out from 
the integral in \eq{eq:sigmasumq}. 
Although numerical effects related to 
adopting this approximation remain small, 
in our numerical study we use the more accurate expression \eq{eq:Msq0}, 
in which case the matrix element carries a dependence on $k_A$
and must be kept under the integral.

\section{Electron momentum density}
The key input for the calculation of the resonant cross-section is the electron momentum distribution $n_A (k_A)$.  For most of the elements this quantity has been extracted 
directly from measurements of the CP~\cite{Cooper_1985}.  
In this Letter, we focus on electron momentum density from spherically averaged Compton profiles, $J(p)$, defined as~\cite{2017PhyB..521..361A}
\be\label{eq:CP}
J(p) = \frac{1}{2} \int_{\left|p\right|}^\infty \rho(k) k dk,
\ee
where $\rho(k)$ is the electron momentum distribution normalised as  
$\int_{-\infty}^\infty J(p)dp= Z$ which, in our notation, can be rewritten as $\rho(k)=\frac{n(k)}{2\pi^2}$. The  electron momentum density distribution and the CP 
are then related as: 
\be\label{eq:nJ}
n(k) = -\frac{(2\pi)^2}{k}\frac{dJ(k)}{dk}.
\ee
 
\begin{figure*}[ht!]
    \centering
    \includegraphics[width=0.49 \textwidth]{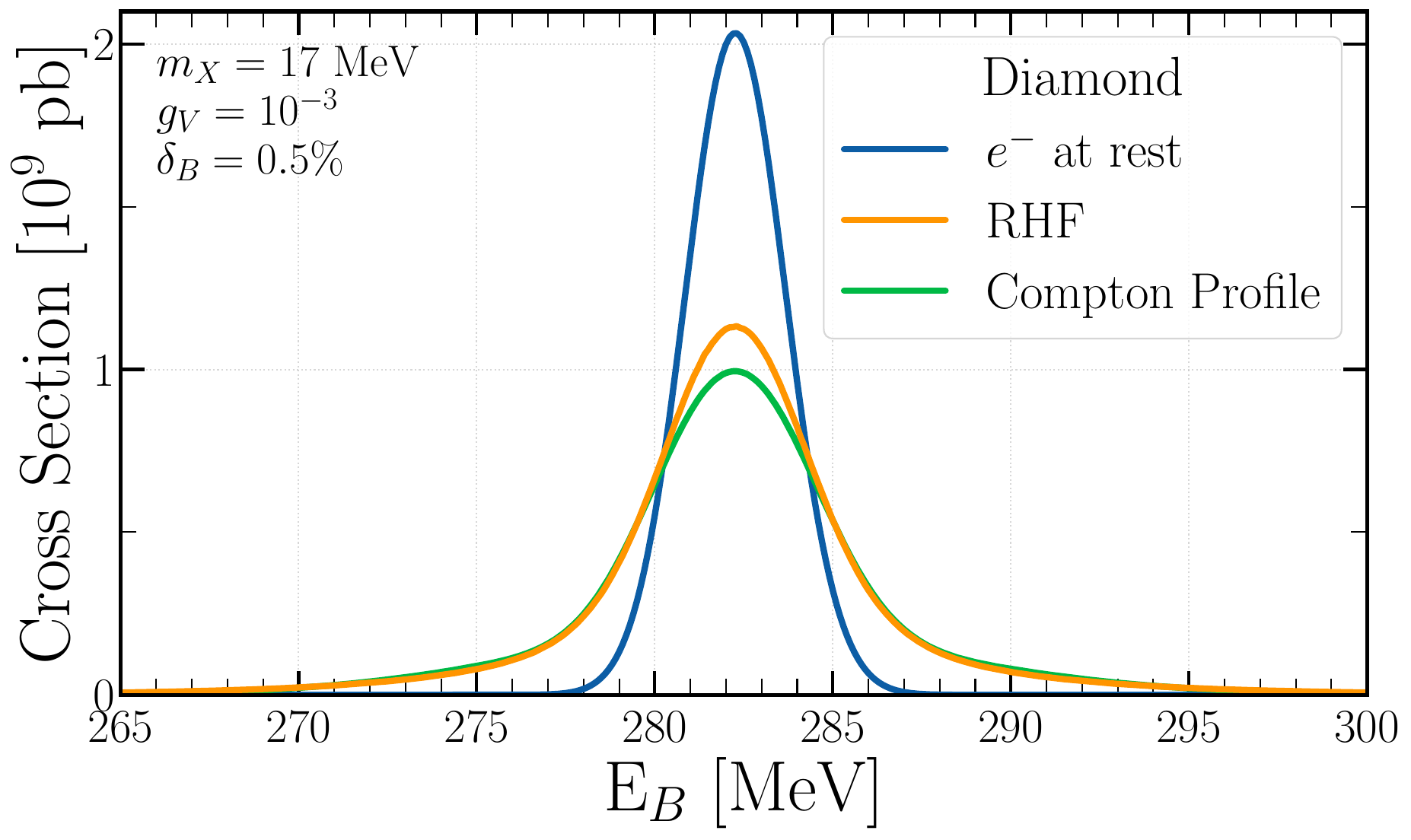}
    \includegraphics[width=0.49 \textwidth]{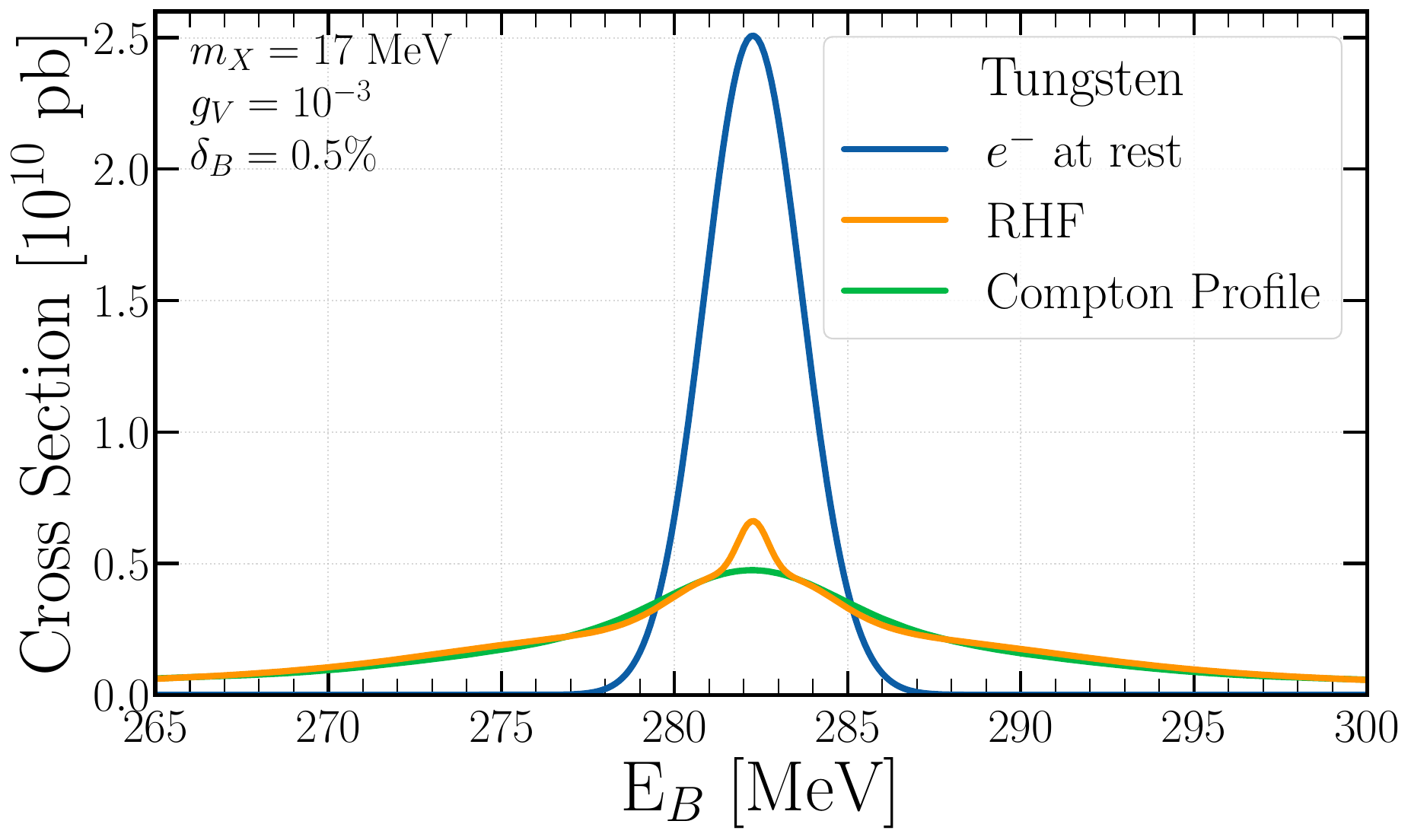}
    \caption{
    Cross section for resonant production  of a  new vector boson
    with $m_X=17\,$MeV and $g_V = 10^{-3}$, including the 
    effects of positron beam energy spread. 
    The blue curve assumes electrons at rest. The orange curve is obtained with RHF wave functions. The green curve is obtained by using the diamond (left plot) and tungsten (right plot) CP.  
    }
    \label{fig:cs}
\end{figure*}

The electron momentum distribution of materials  can also be obtained from {\it  ab initio} theoretical calculations. Approximate expressions can be derived 
by using Roothan-Hartree-Fock (RHF) wave functions~\cite{huzinaga1965gaussian,steinborn1973rotation,clementi1974roothaan,kaijser1977evaluation,filter1978extremely,maretis1979talmi,weniger1983fourier,weniger1985weakly}, see Appendix~\ref{app:nA}.  

In the next section we will focus on the PADME experiment~\cite{Raggi:2014zpa,Raggi:2015gza} that is using a polycrystalline diamond $100\,\mu$m target, for which CP data from experiment~\cite{PhysRev.176.900} 
as well as refined theoretical estimates~\cite{2017PhyB..521..361A} are available.
 We extend the $k_A$ range of these CP by using the RHF wave functions method. 
 The electronic structure of carbon is $1s^22s^22p^2$. However, in the diamond crystal structure one electron is promoted from  $2s$ to the $2p$ orbital to increase the covalent bounds.
 $2s 2p^3$ electrons of one atom then undergo $sp^3$ hybridization bonding it to four other  atoms.
RHF wave functions can  be used to perform $sp^3$ hybridization, as detailed in Appendix~\ref{app:nA}. 

Besides carbon, we also study the case of a high $Z$ tungsten target,  that has the  electronic structure  $[\mathrm{Xe}]\, 6s^24f^{14}5d^4$.
We use the CP for tungsten from Table~I of Ref.~\cite{PhysRevB.38.12208}, and 
 for momenta larger than $p=7\,\mathrm{a.u.}\simeq 27\,$keV up to 
$p=100\,\mathrm{a.u.}\simeq 370\,$keV we complement those data with the theoretical CP 
derived in the Dirac Hartree-Fock formalism given in Ref.~\cite{BIGGS1975201}. For even larger momenta we use 
the code DBSR-HF~\cite{ZATSARINNY2016287} to numerically estimate the contribution of the core orbitals up to the $4s$ shell.\footnote{See Appendix~\ref{app:dbsr} for more details.}

\section{Searches for light new bosons } 
Our primary goal is to assess  the impact of the motion of atomic electrons 
on the production of light new bosons via $e^+e^-$ resonant annihilation. 
In computing the cross section one must also take into account the energy distribution of positrons in the beam, that can be described by a Gaussian $\mathcal{G}(E, E_B, \sigma_B)$ centered at $E_B$ and with standard deviation $\sigma_B$:
\begin{equation}
\label{eq:beamsmearing}
\sigma_\mathrm{final}(E_B,\sigma_B) = \int dE\ \mathcal{G}(E, E_B, \sigma_B) \sigma(E).
\end{equation} 
In all our computations we have assumed  a beam spread $\delta_B\equiv \sigma_B/E_B=0.5\%$.

We will now focus on the interesting case of the $X_{17}$ boson assuming $m_X=17$ MeV. However, we stress that the effects that we will illustrate do  not depend on the nature or mass of the new particle.  Figure~\ref{fig:cs} shows a comparison between  cross sections evaluated with different assumptions as a function of the  beam energy. 
Results for a diamond target are given in the left panel, where the blue curve is obtained by assuming  electrons at rest, so that the spread is entirely due to $\sigma_B$. The orange curve shows the cross section using the RHF approximation, while the green curve is obtained with the CP from~\cite{2017PhyB..521..361A}. These last two curves are more spread due to the motion of the atomic electrons. The peculiar structure of the green curve is due to the fact that the two core electrons contribute to the broad tails, while the four bond electrons mainly contribute to the central peak. 
In the right panel we show the same results for tungsten. In this case the  cross section exhibits a more significant energy spreading. This is due to the wide 
range of velocities from high-momentum core electrons.
In both cases, the corrected cross-section is starkly different from the electron-at-rest approximation. 
Clearly, the smearing of the resonance will have an important  impact on 
searches for resonance peaks at fixed target experiments. 
 \begin{figure*}[ht!]
    \centering
    \includegraphics[width=0.49 \textwidth]{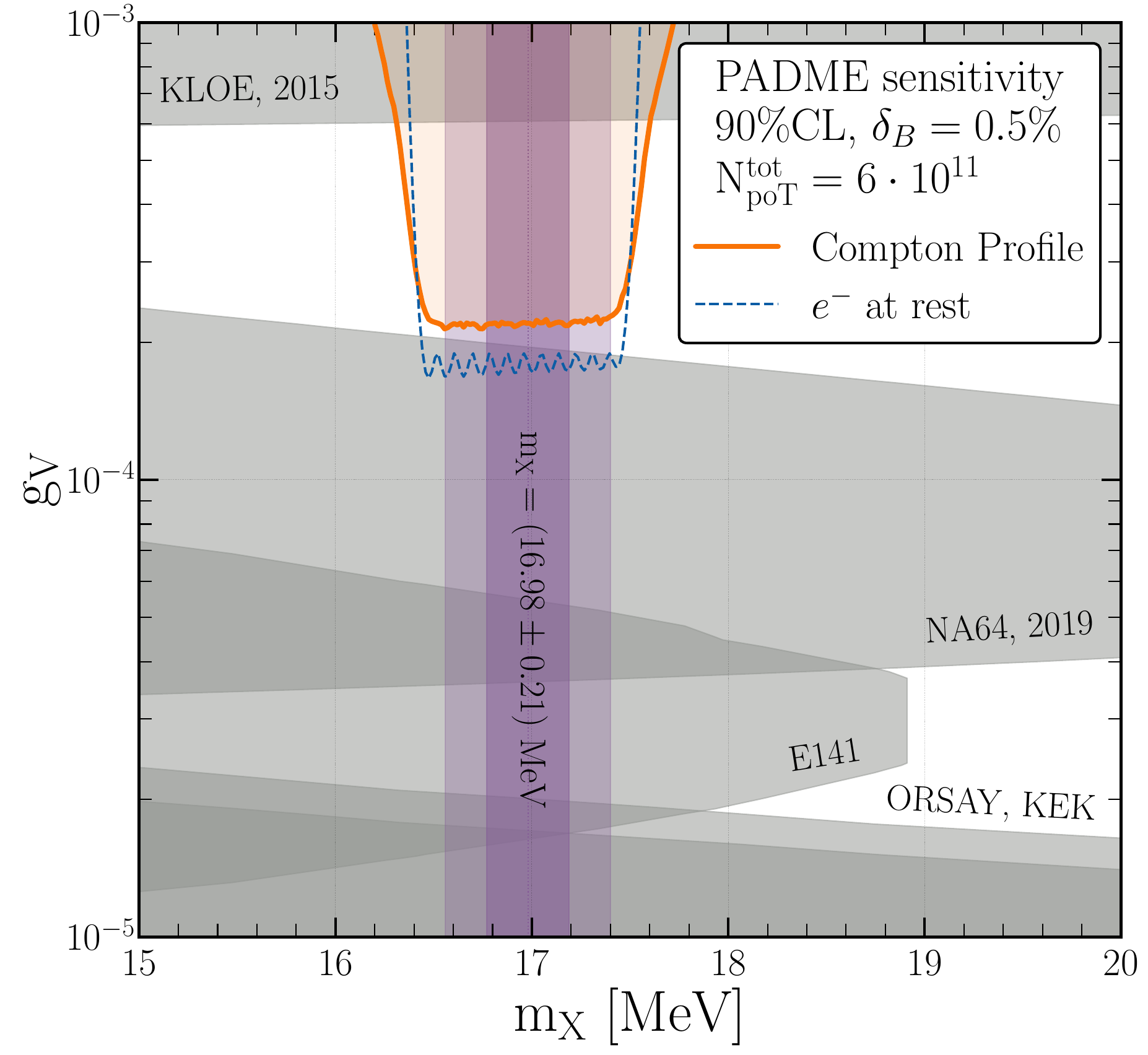}
    \includegraphics[width=0.49 \textwidth]{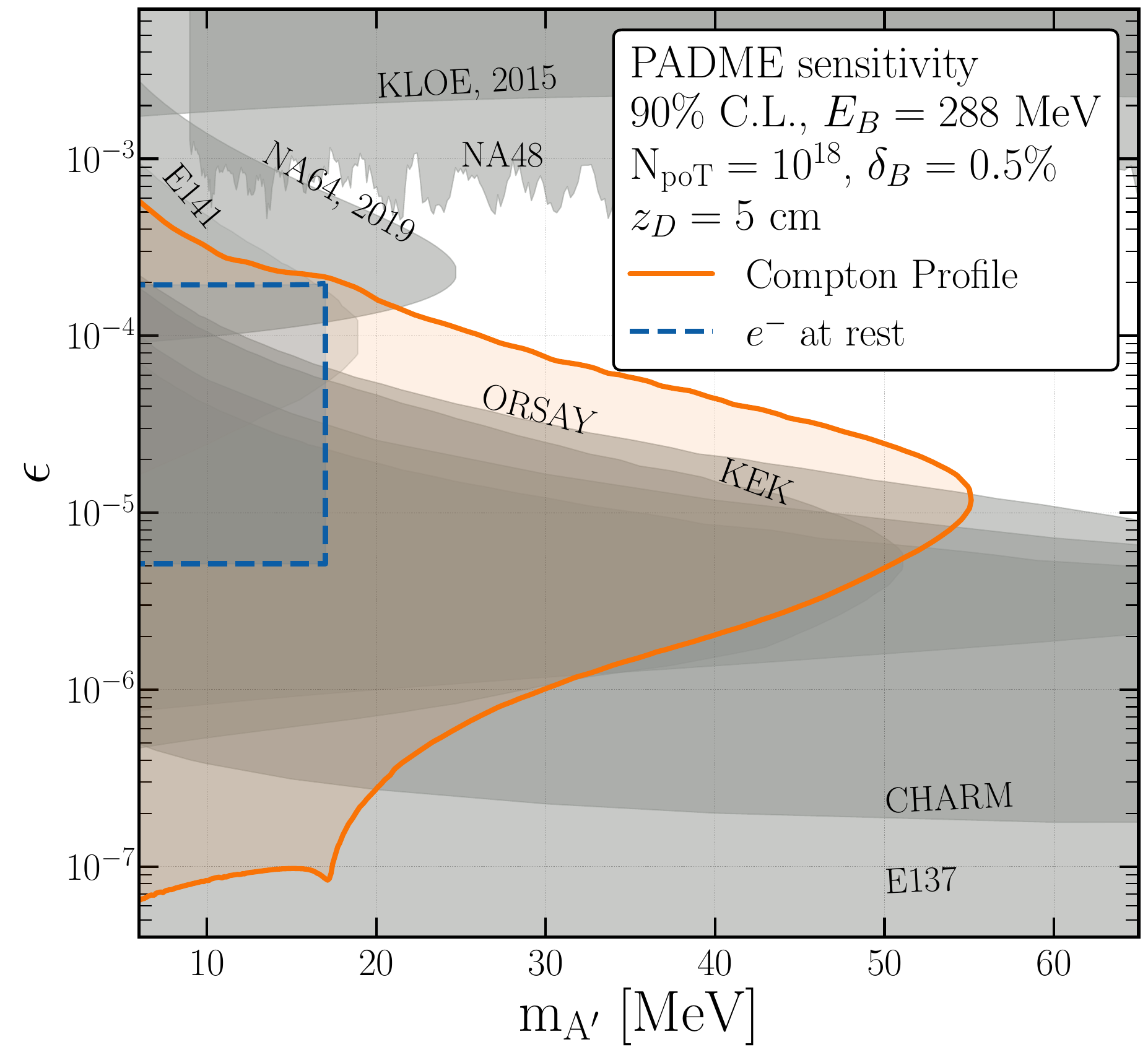}
    \caption{\textbf{Left:}
    Projected 90\% C.L. sensitivity of PADME run-III on $g_V$ as a function of the $X_{17}$ mass assuming atomic electrons at rest (dashed blue) and including  electron velocity effects using the diamond CP (solid orange).
    The dark (light) violet region represents the $m_X$ range from the combined ATOMKI result at 68\%   (95\%)~C.L.~\cite{Krasznahorkay:2015iga,Krasznahorkay:2018snd,Krasznahorkay:2021joi}.  
    \textbf{Right:} Projected 90\% C.L. sensitivity on the DP kinetic mixing parameter $\epsilon$ as a function of the DP mass for positrons impinging on a $5\,$cm thick tungsten target. The  
    dashed blue line assumes electron at rest.  The dotted orange line 
    includes electron velocity effects  using the tungsten CP.  
    In both plots, the gray shaded regions represent excluded regions (see text).
  }
    \label{fig:sensitivity}
\end{figure*}

In Fig.~\ref{fig:sensitivity} we show the impact of electron velocities 
on PADME searches for resonantly produced vector bosons. 
The PADME sensitivity to $X_{17}$ searches using a thin diamond target is depicted in the left panel, where the reach in the coupling $g_V$ is plotted  as a function of $m_X$.
A scan with 12  energy bins covering the  range $E_B = [265,\,297]\,$MeV 
with a total of $6\cdot 10^{11}$ positrons on target and assuming a $0.5\%$ energy spread,
which corresponds to the nominal parameters for the PADME run-III search, has been assumed.    
The shaded violet regions show the $1\sigma$ and $2\sigma$ range for $m_X\sim 16.98\pm0.21\,$MeV 
from combining the statistical errors from the different ATOMKI measurements~\cite{Krasznahorkay:2015iga,Krasznahorkay:2018snd,Krasznahorkay:2021joi}
and adding in quadrature a common systematic error of $0.20\,$MeV. 
The gray shaded regions are excluded by KLOE~\cite{Anastasi:2015qla}, E141~\cite{Riordan:1987aw}, NA64~\cite{NA64:2021aiq}, ORSAY~\cite{Davier:1989wz} and KEK~\cite{Konaka:1986cb}.

The orange line in the left panel shows the projected sensitivity with electron 
velocity effects included via the CP. 
The dashed blue line shows, for comparison, the results previously obtained 
by assuming electrons at rest~\cite{Darme:2022zfw}. 
Both these lines have been obtained by 
assuming a negligible $\gamma\gamma$ background, while $7.5\times 10^4$ background events 
are expected from $s$- and $t$-channel $e^+e^-\to e^+e^-$. 
The acceptance was evaluated by requiring that the energy of the final states $e^\pm$ 
originating from $X_{17}$ decays 
satisfies $E_\pm>100$ MeV and the azimuthal angle is in the range $25.5\lesssim \theta_{\pm}/\mathrm{mrad}\lesssim 77$, leading to an acceptance of $20\%$~\cite{Darme:2022zfw}.  Systematic uncertainties have been neglected.
The plot shows a certain loss in sensitivity once the effect of electron velocities is included. This occurs because the signal gets distributed over a broader range of energies, resulting in a reduction of the signal-to-noise ratio.

In the right panel of Fig.~\ref{fig:sensitivity} the 
grey regions represent the exclusion  limits for DP searches with 
$A'$ decaying  into $e^+e^-$ pairs with unit branching ratio, from  the KLOE~\cite{Anastasi:2015qla}, NA64~\cite{NA64:2020xxh}, ORSAY~\cite{Davier:1989wz}, KEK~\cite{Konaka:1986cb}, E137~\cite{Andreas:2012mt,Marsicano:2018krp},  CHARM~\cite{Tsai:2019buq} and SLAC E141~\cite{Riordan:1987aw} experiments.
The orange  region depicts the forecasted $90\%$ C.L. sensitivity (corresponding to $\sim2.7$ signal events) achievable with a positron-beam with $E_B = 288$ MeV 
impinging on a $5\,$cm-thick tungsten  target, assuming a total of $10^{18}$ poT, 
in the background-free limit. 
We see that the contribution of  electrons with large momenta allows to probe mass regions 
that extend to values much larger than one would find assuming electron-at-rest. In fact, practically all of the region from DP masses around $25$ MeV up to $80$ MeV
can be probed thanks to the large  tail of the electron momentum distribution. 
The blue region represents the sensitivity to the $X_{17}$ under the same experimental conditions, 
but assuming electrons at rest. 


\section{Outlook and conclusion} 
In this work we have discussed a prescription for including the effects of 
atomic electron momenta in evaluating the cross section for 
resonant positron annihilation on fixed targets. 
The electron momentum density can be obtained  both from the CP of the target material, 
or directly from theoretical computation. We have argued that even a relatively simple approach as  
using RHF wave functions largely improves on the electron-at-rest approximation, 
as is shown in Fig.~\ref{fig:cs}. 
We have studied the relevance of these effects in a low $Z$ material (crystalline carbon) 
as well as in a high $Z$ material, finding in both cases 
stark differences with the electron-at-rest cross sections.  
This implies that, in order to obtain reliable quantitative predictions,   
experiments planning to  search for new bosons via their resonant production 
in fixed targets must necessarily  account for atomic electron velocity effects. 

This work paves the way for new search strategies where high-$Z$ targets can be leveraged  
to expand the  mass region that can be probed for dark bosons, even when the beam energy 
is held constant. Indeed, our key finding  is that high-$Z$ atoms can emulate  particle physics accelerators by supplying a non-negligible density of  high momentum electrons that, when colliding head-on with beam positrons, will yield a large increase in the CoM energy.  

While we have focused on diamond and tungsten targets, for which the 
electron momentum distribution can be taken as approximately isotropic, 
our result, Eq.~\eqref{eq:dCScrysFinal}, can be applied also to non-isotropic materials. 
This is important because targets characterized by significant anisotropies may 
give rise to observable effects on resonant production rates when the orientation 
is changed. Since non-resonant background processes have a much weaker dependence on the 
electron momentum distribution, it can be speculated that by 
comparing data for different orientations of an anisotropic target  
might help to separate signal from background events. 
We leave a study of this possibility for future work.
\\
\begin{acknowledgments}
\noindent
We thank the authors of Ref~~\cite{Plestid:2024xzh} 
and in particular R. Plestid for contributing to identify 
an error in the first version of this paper.
We thank P. Gianotti, M. Raggi, T. Spadaro, P. Valente, and all the members of the PADME collaboration for several useful exchanges regarding the PADME experiment. F.A.A., G.G.d.C. and E.N. were supported by the INFN ``Iniziativa Specifica" Theoretical Astroparticle Physics (TAsP), with F.A.A. receiving additional support from an INFN Cabibbo Fellowship, call 2022. L.D. has been supported by the European Union’s Horizon 2020 research and innovation programme under the Marie Skłodowska-Curie grant agreement No 101028626 from 01.09.2021 to 31.08.2023.  G.G.d.C. acknowledges LNF and Sapienza University for hospitality at various stages of this work. The work of E.N. was supported  by the Estonian Research Council grant PRG1884. We acknowledge support from the CoE grant TK202 “Foundations of the Universe” and from the CERN and ESA Science Consortium of Estonia, grants RVTT3 and RVTT7. We also acknowledge support by COST (European Co- operation in Science and Technology) via the COST Action COSMIC WISPers CA21106. 
\end{acknowledgments}

\newpage

\appendix
\onecolumngrid

\section{Cross section for resonant production}
\label{app:CS}

The differential cross section can be written as:
\begin{eqnarray}
 d\sigma&=&\frac{d^3p}{(2\pi)^3}\frac{1}{2E_X}\int d^2b\int \frac{d^3k_A}{(2\pi)^3}\frac{d^3\Bar{k}_A}{(2\pi)^3}\int \frac{d^3k_B}{(2\pi)^3}\frac{d^3\bar{k}_B}{(2\pi)^3}\,%
\frac{\phi_A(\bm{k}_A)\phi_A^*(\bm{\bar{k}}_A)}{\sqrt{2E_{k_A}2{E}_{\bar{k}_A}}} \frac{\phi_B(\bm{k}_B)\phi_B^*(\bm{\bar{k}}_B)}{\sqrt{2E_B2\bar{E}_B}} e^{-i\bm{b}\cdot(\bm{k}_B-\bm{\bar{k}}_B)}\nonumber\\
 &&\times \prescript{}{out}{\left<\bm{p}|\bm{k}_A,\bm{k}_B\right>_{in}}\prescript{}{out}{\left<\bm{p}|\bm{\bar{k}}_A,\bm{\bar{k}}_B\right>_{in}^*},
\end{eqnarray}
where $\bm{b}$ is the impact parameter vector laying in a plane transverse to the beam direction, 
$\phi(\bm{k}_A)\ (\phi(\bm{k}_B)),\ \bm{k}_A\ (\bm{k}_B)$ and ${E_{k_A}}\ (E_B)$ represent the wave function, momentum and energy of the free electron (positron), satisfying $E^2=m^2+k^2$, while $p$ and $E_X$ stand for the final particle momentum and energy. Barred variables have the same meaning as non-barred ones. The integral over the impact factor is straightforward and yields a factor $(2\pi)^2\delta^{(2)}(\bm{k}_B^{\perp}-\bm{\bar{k}}_B^{\perp})$, where the superindex  $\perp$ refers to vectors lying on the plane orthogonal to the beam. The overlap of the initial and final states can be written using the matrix element and energy-momentum conservation (the same applies to the $\bar{k}_A$ and $\bar{k}_B$ momenta), i.e.:
\be
\prescript{}{out}{\left<\bm{p}|\bm{k}_A,\bm{k}_B\right>_{in}}=i(2\pi)^4\delta\left(E_A+E_B-E_X\right)\delta^{(3)}\left(\bm{k_A}+\bm{k_B}-\bm{p}\right)\mathcal{M},
\ee
where $\mathcal{M}$ is the amplitude for the free process and $E_A=m_e$ stands for the energy of the bound electron (neglecting the binding energy). Six of the Dirac deltas involving the barred variables can be used to perform the six integrals in the barred three-momenta. As a result, we have:
\begin{equation}
d\sigma =\frac{d^3p}{(2\pi)^3} \, \int\frac{d^3k_A}{(2\pi)^3}\int \frac{d^3k_B}{(2\pi)^3} \,\frac{(2\pi)^4\delta\left(E_A+E_B-E_X\right)\delta^{(3)}\left(\bm{k_A}+\bm{k_B}-\bm{p}\right)}{2E_X2E_{k_A}2E_B\left|v_A-v_B\right|} \left|\phi_A(\bm{k}_A)\right|^2\left|\mathcal{M}(k_A,k_B\rightarrow p)\right|^2\left|\phi_B(\bm{k}_B)\right|^2,
\end{equation}
where  $E_{k_A} \equiv \sqrt{\bm{k}_A^2 + m_e^2}\,$.
Since the positrons in the beam correspond to plane waves 
with a well defined momentum
$\bm{p}_B$, their wave function satisfies
\be
\int \frac{dk_B^3}{(2\pi)^3} \left|\phi_B(k_B)\right|^2 = 1,\,\,\qquad \left|\phi_B(k_B)\right|^2 = (2\pi)^3 \delta^{(3)}(\bm{p}_B-\bm{k}_B),
\label{eq:conditionfree}
\ee
which can be used to perform the integral over $\bm{k}_B$. We find
\begin{equation}
d\sigma =\frac{d^3p}{(2\pi)^3} \int\frac{d^3k_A}{(2\pi)^3}  \frac{(2\pi)^4n(\bm{k}_A)\left|\mathcal{M}(k_A,p_B\rightarrow p)\right|^2 }{8 E_X \left|E_Bk_A^z-E_{k_A}p_B^z\right|}\delta\left(E_A+E_B-E_X\right)\delta^{(3)}\left(\bm{k_A}+\bm{p_B}-\bm{p}\right)),
\label{eq:cs}
\end{equation}
where we used $\left|v_A-v_B\right|E_{k_A}E_B=\left|E_Bk_A^z-E_{k_A}p_B^z\right|$. 
In Eq.~\eqref{eq:cs}, we have introduced the electron momentum density function $n(\bm{k}_A) =\sum_{n,l} \left|\phi_{n,l}(\bm{k}_A)\right|^2$, normalized to the total number of electrons in the atom $\int\frac{ d^3k_A}{{(2\pi)^3}} \; n(\bm{k}_A) =  Z$.
Integrating over $d^3p$ yields
\be
\sigma = \int d^3 k_A\, \frac{\left|\mathcal{M}\right|^2n(\mathbf{k}_A)\,\delta(E_A +E_B-E_X(\bm{k}_A))}{32\pi^2E_X(\mathbf{k}_A)|E_B k_A^z - E_{k_A} p_B^z|}\,,
\ee
where 
$E_X=\sqrt{k_A^2+p_B^2+2k_Ap_B x+m_X^2}$, 
$x=\cos\theta$, with $ \theta$ the angle  between $\bm{p}_B$ and $\bm{k}_A$ 
and, neglecting atomic binding energies, $E_A= m_e$.

In the case of an isotropic electron momentum distribution $n(\bm{k}_A)=n(k_A)$, after integrating in the azimuthal angle $\phi$ and introducing the functions $F(x)$ and $x_0(k_A)$ to account for the solution of the energy conservation condition
\begin{equation}\label{eq:x0}
F(k_A,x)=\frac{\partial}{\partial x}(E_A+E_B-E_X)=-\frac{k_A p_B}{E_X(k_A,x)}\,,\qquad
x_0(k_A) = \frac{2E_A E_B+2m_e^2-m_X^2-k_A^2 }{2k_Ap_B},
\end{equation} 
we obtain 
\be
\label{eq:cscrystal}
\sigma = \int_{k_A^{\mathrm{min}}}^{k_A^\mathrm{max}}dk_A\frac{\left|\mathcal{M}\right|^2k_A n(k_A) }{16\pi p_B|E_B k_A x_0(k_A)-E_{k_A} p_B|}.
\ee
Energy conservation implies  $-1\leq x_0 \leq 1$, or equivalently
\begin{eqnarray}
    k_A^\mathrm{max,min} = \left|p_B \pm \sqrt{ \left(E_A + E_B \right)^2-m_X^2} \right|\, \nonumber \,.
\end{eqnarray}
The matrix element can be written generically in terms of products of four momenta as follows:
\be
\left|\cM_{free}\right|^2=g_V^2\left(3m_e^2+K_A\cdot P_B+\frac{2}{m_X^2}K_A\cdot P_X\ P_B\cdot P_X\right),
\label{eq:Msq}
\ee
where the four momenta, denoted by capital letters,
correspond to free one particle states. In particular,  
$K_A$ has time-like component  $E_{k_A}=\sqrt{k_A^2+m^2}$. 
The matrix element remains within the integral, 
which implies that the three-momentum and energy delta functions operate on it. The first step, three-momentum conservation, implies the following for the four-vector products:
\bg
K_A\cdot P_B = E_{k_A}E_B-\bm{k_A}\cdot \bm{p_B} = E_{k_A}E_B-k_A p_B \cos{\theta},\\
K_A\cdot P_X = E_{k_A}E_X-\bm{k_A}\cdot \bm{p_X} =E_{k_A}E_X-\bm{k_A}\cdot \left(\bm{k_A}+\bm{p_B}\right) = E_{k_A}E_X-k_A^2-k_A p_B \cos{\theta},\\
P_B\cdot P_X =  E_{B}E_X-\bm{p_B}\cdot \bm{p_X} =E_{B}E_X-\bm{p_B}\cdot \left(\bm{k_A}+\bm{p_B}\right) = E_{B}E_X-p_B^2-k_A p_B \cos{\theta}.
\eg
Then, integrating the delta for energy conservation implies $\cos{\theta}=x_0\left(k_A\right)$, so that the final expression for the matrix element reads
\ba
\left|\cM_{free}\right|^2=g_V^2\Big[3&m_e^2+ E_{k_A}E_B-k_A p_B x_0\left(k_A\right)\\
+&\frac{2}{m_X^2}\left(E_B E_X-p_B\left(p_B+k_Ax_0\left(k_A\right)\right)\right)\left(E_{k_A} E_X-k_A\left(k_A+p_Bx_0\left(k_A\right)\right)\right)\Big],
\ea
with $x_0\left(k_A\right)$ as defined in Eq.~\eqref{eq:x0}
and $E_X=m_e+E_B$ given by energy conservation.
In the limit $E_B \gg m_e,E_{k_A}$ this simplifies to 
\ba
\left|\cM_{free}\right|^2 \simeq g_V^2 \left[m_X^2+ 2 m_e^2 + E_B (E_{k_A} - m_e)\, \left( 2 + \frac{k_A^2}{m_X^2} \right)\right].
\ea

Finally, note that the limit 
of free electrons at rest can be recovered from \eq{eq:cscrystal} 
by replacing $n_A$ with a delta-function (as in Eq.~\eqref{eq:conditionfree}) multiplied by $Z$.

\section{Electron momentum density distribution in diamond}
\label{app:nA}
The most important quantity in the result in~\eqref{eq:cscrystal} is the electron momentum density function $n_A (k_A)$.

\paragraph{Roothan-Hartree-Fock (RHF)  wave functions.}

\noindent
The RHF equations describe the atomic wave functions, characterized by the principal, azimuthal and magnetic quantum numbers $n$, $\ell$ and $m$, by separating the radial and angular part. The angular part is described by the spherical harmonics (in the Condon-Shortley phase convention~\cite{condon1935theory}) $Y_\ell^m(\hat{x})$, while the radial one is described in terms of Slater-type orbitals (STO)~\cite{Slater:1930zz}:
\be
   R(r,Z,n) = a_0^{-3/2} \frac{(2 Z)^{n+1/2}}{\sqrt{(2n)!}} \left ( \frac{r}{a_0} \right)^{n-1} e^{-\frac{Z r}{a_0}},
\ee
with $a_0$ the Bohr radius and $Z$ the effective nuclear charge. The  atomic wave function can then be written as:
\be
  \psi_{n \ell m}^{\mathrm{atom}}(\vec{x}) = \sum_j C_{j\ell n} R(r, Z_{j\ell}, n_{j\ell}) Y_\ell^m(\hat{x}),
  \label{eq:WF}
\ee
where $C_{j\ell n}$, $Z_{j\ell}$ and $n_{j\ell}$ are tabulated e.g. in Refs.~~\cite{Bunge:1993jsz,McLean:1981mjg}. The Fourier transform of this wave function can be computed analytically~\cite{Belkic_1989}. 

\paragraph{Fourier transform of the wave function.}

\noindent 
The Fourier transform of the RHF STO wave function in equation~\eqref{eq:WF} can be written analytically~\cite{Belkic_1989}:
\begin{eqnarray}
     \tilde{\psi}_{n \ell m}^{\mathrm{atom}}(\vec{q}) &=& \sum_j C_{j\ell n} \int \frac{d^3x}{(2\pi)^3}\,e^{i\vec{q}\cdot \vec{x}} R(x,Z_{j\ell},n_{j\ell}) Y_\ell^m(\hat{x})
     = \sum_j C_{j\ell n} \,\tilde{\chi}(q,Z_{j\ell},n_{j\ell},\ell) Y_\ell^m(\hat{q})\nonumber\\
    \tilde{\chi}(q,Z_{j\ell},n_{j\ell},\ell) &=& a_0^{3/2}\frac{2^{n_{j\ell}-1}(n_{j\ell}-\ell)!}{\pi^2} \, \left({i a_0 q} \right)^{\ell} Z_{j\ell}^{n_{j\ell}-\ell} 
    \frac{(2Z_{j\ell})^{n_{j\ell}+1/2}}{\sqrt{(2n_{j\ell})!}} \, \sum_{s=0}^{\lfloor (n_{j\ell}-\ell)/2\rfloor} \frac{\omega_s^{n_{j\ell}\ell}}{((a_0 q)^2 +Z_{j\ell}^2)^{n_{j\ell}-s+1}},\nonumber\\
    \omega_s^{n_{j\ell}\ell} &=& (-4Z_{j\ell}^2)^{-s} \frac{(n_{j\ell}-s)!}{s! (n_{j\ell}-\ell-2s)!},
    \label{eq:RHF}
\end{eqnarray} 
where the upper limit $ \lfloor (n_{j\ell}-\ell)/2\rfloor$  in the summation denotes the floor of $(n_{j\ell}-\ell)/2$, i.e., the largest integer less or equal to the argument, and  the coefficients $C_{j\ell n}$, $n_{j\ell}$ and $Z_{j\ell}$ can be found in Ref.~\cite{Bunge:1993jsz,McLean:1981mjg}.

\paragraph{Hybridization for diamond.}
\label{app:sp3}
\noindent
The electronic structure of carbon in diamond is $1s^22s^1 2p^3$, with the valence electrons exhibiting $sp^3$ hybridization. We can then construct the electron wave functions for the $n=2$ electrons:
\begin{eqnarray}
\psi^\mathrm{atom}_{sp3a} &=& \tilde{\psi}_{20(0)}^{\mathrm{atom}} +\tilde{\psi}_{21(+1)}^{\mathrm{atom}}+\tilde{\psi}_{21(0)}^{\mathrm{atom}}+\tilde{\psi}_{21(-1)}^{\mathrm{atom}}\,,\nonumber\\
\psi^\mathrm{atom}_{sp3b} &=& \tilde{\psi}_{20(0)}^{\mathrm{atom}} +\tilde{\psi}_{21(+1)}^{\mathrm{atom}}-\tilde{\psi}_{21(0)}^{\mathrm{atom}}-\tilde{\psi}_{21(-1)}^{\mathrm{atom}}\,,\nonumber\\
\psi^\mathrm{atom}_{sp3c} &=& \tilde{\psi}_{20(0)}^{\mathrm{atom}} -\tilde{\psi}_{21(+1)}^{\mathrm{atom}}-\tilde{\psi}_{21(0)}^{\mathrm{atom}}+\tilde{\psi}_{21(-1)}^{\mathrm{atom}}\,,\nonumber\\
\psi^\mathrm{atom}_{sp3d} &=& \tilde{\psi}_{20(0)}^{\mathrm{atom}} -\tilde{\psi}_{21(+1)}^{\mathrm{atom}}+\tilde{\psi}_{21(0)}^{\mathrm{atom}}-\tilde{\psi}_{21(-1)}^{\mathrm{atom}}.\nonumber
\end{eqnarray}
On the other hand, the $1s$ electrons remain in their RHF STO form defined in equation~\eqref{eq:RHF}.

\section{Electron momentum density distribution in large-Z materials}
\label{app:dbsr}

For large-Z material, the core electrons are typically relativistic and the study of the large momentum limit should thus rely on relativistic quantum mechanical methods. Fortunately, several state-of-the art numerical codes are available. We have used  codes based on the Dirac-Hartree-Fock (DHF) method. In the case of a Coulomb-like potential (and thus a point-like nuclei), the results for the Dirac wave function are well-known and can be found in various textbooks (see e.g.~\cite{Bethe:1957ncq} chap. 14).  %

\begin{figure*}[ht!]
    \centering
    \includegraphics[width=0.6 \textwidth]{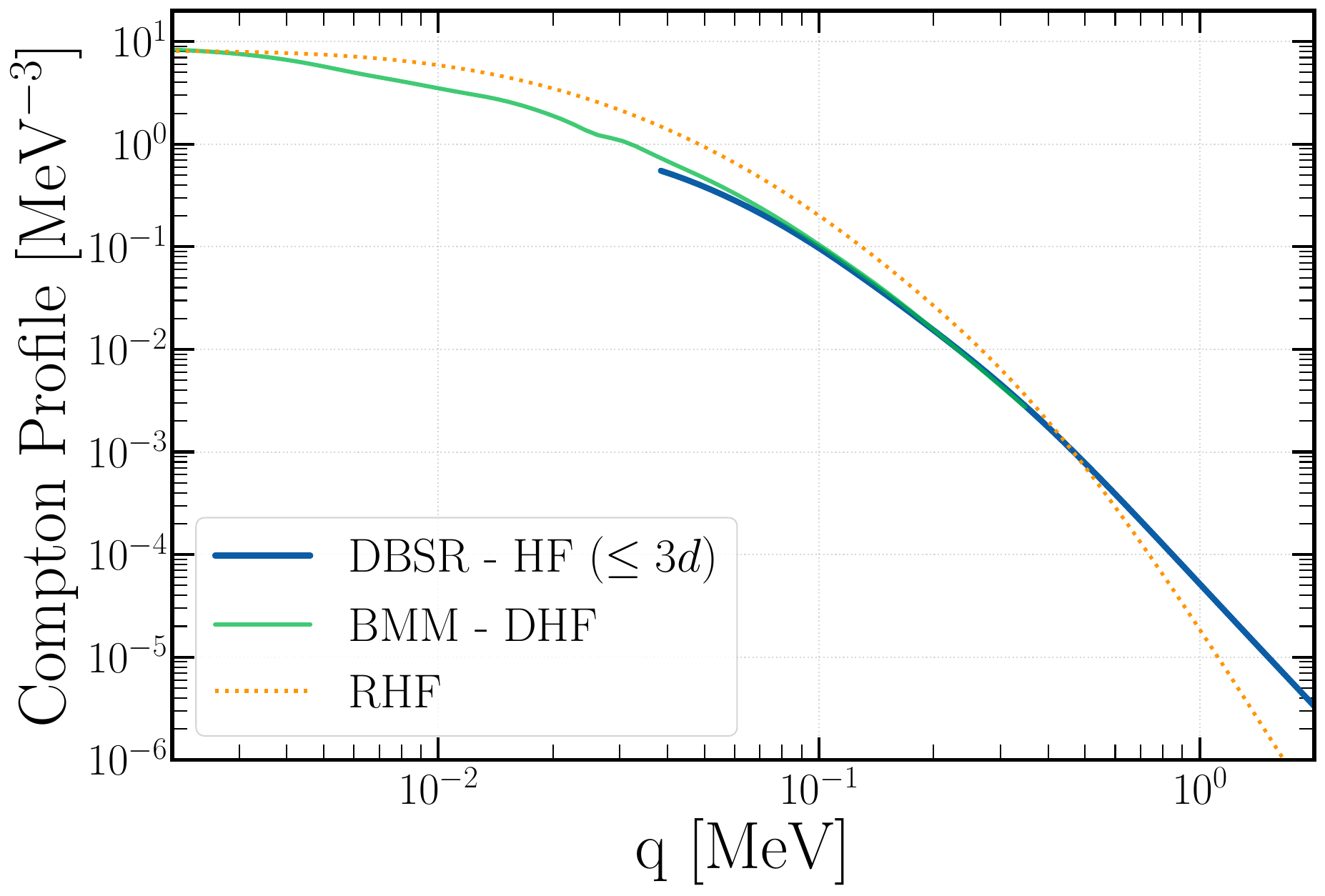}
    \caption{Matching between the different approaches to the electron density distribution at large momentum, represented in term of the isotropic Compton profile $J(q)$. The green line is the DHF as extracted from~\cite{BIGGS1975201}, the blue line the result from DBSR-HF~\cite{ZATSARINNY2016287}. We show for comparison the non-relativistic HF from~\cite{McLean:1981mjg} in dotted orange.}
    \label{fig:matching}
\end{figure*}

Various improvements have been made throughout the years, in particular with a better description of the charge density inside the nucleus, various interaction terms between electrons and a modelisation of QED corrections (for a recent review, see for instance Ref.~\cite{smits:hal-03884148}). We rely on the implementation of these effects in the codes GRASP18~\cite{JONSSON20132197,2019CoPhC.237..184F} and DBSR-HF~\cite{ZATSARINNY2016287}. More precisely, these codes are used to find the ``large'' and ``little'' radial wave functions components (resp. $P(r)$ and $Q(r)$) for each orbital. The momentum density distribution is then obtained as:
\begin{equation}
\rho_q(k) = N_q (|\gamma_{q,P}|^2 + |\gamma_{q,Q}|^2)
\end{equation}
 where  $N_q$ is the number of electrons in each orbital and $q$ refers to the DHF orbitals. Note that the latter are  split with respect to the non-relativistic case due to the spin-orbit contributions linking the orbital moment $\ell$ and electron spin.\footnote{We use the same notation as in the outputs of the above atomic physics codes. For instance, there are $2$ states in the s-shells, $2$ in $p-$ shells, $4$ in $p$ shells, $4$ in $d-$ shells and $6$ in $d$ shells.} We also introduced $\gamma_{q,P}$ and $\gamma_{q,Q}$ which are integrals over the radius $r$. For the orbitals $s$ and $p$ we have:
\begin{equation}
\gamma_P(k) = \sqrt{\frac{2}{\pi}} \int_{0}^\infty  P(r) j_{\ell} (k \, r) dr
\end{equation}
where $\ell=0,1$ for $s,p$ shells and 
$j_{\ell}$ is a spherical Bessel function of the first kind. For $\gamma_Q$, we have:
\begin{align}
\gamma_Q(k) = \sqrt{\frac{2}{\pi}}  \int_{0}^\infty  Q(r) j_{\ell_i} (k \, r) dr \ \quad \textrm{ with } \ell_i = 
\begin{cases}
    1 \textrm{ for s shells}\\
    0 \textrm{ for p$-$ shells}\\
    2 \textrm{ for p shells}
\end{cases}
\end{align}
The $d$ and higher shells behave as the $p$ shell, with e.g. $\ell_i  = \ell - 1 = 1$ for the $d-$ and  $\ell_i  = \ell + 1 = 3$ for the $d$.
We have cross-checked that the output of both GRASP18 and DBSR-HF  (being based on the same theoretical foundations) agree to an excellent level in the relevant large momentum regime. We further illustrate the matching between the different approaches in Fig.~\ref{fig:matching}, where we have represented the Compton profiles as derived  from~\cite{BIGGS1975201} and from DBSR-HF~\cite{ZATSARINNY2016287}. We find a percent level agreement at momenta around $0.25\,$MeV, where we switch from one to the other.

\section{Effects of electron binding energies}
\label{app:binding}

The result for the 
electron momentum density distribution in diamond
can be easily adapted to account for the binding energy of the electrons. For each shell with binding energy $u_q<0$, the energy  of the electron is given by
\begin{align}
\label{eq:EaBound}
E_{A,q}  = m_e + u_q \,.
\end{align}

For  hydrogen-like atoms, exact expressions for the negative binding energies exist~\cite{Bethe:1957ncq}. For more general cases, they can be evaluated via numerical codes like GRASP~\cite{BIGGS1975201} or DBSR-HF~\cite{ZATSARINNY2016287} or simply from tabulated data (e.g. Ref.~\cite{RevModPhys.39.125}).

The only modification to the expression for the cross section given in the main text is the replacement of  $E_A$ with 
the electron energies  as given in  Eq.~\eqref{eq:EaBound}.
Thus, the main effect of the binding energies is to shift the resonant mass condition, which,
at first order, becomes 
\begin{align}
    E_{res, q} = \frac{m_X^2}{2m_e}\left(1-\frac{u_q}{m_e}\right)-m_e \ .
\end{align}
This amount to a $0.06 \%$ shift for the $1s$ shell of carbon (with $u_{A\,1s} = -300 \,$eV) and a $13 \%$ shift for the $1s$ of tungsten (with $u_{1s} = -69 \,$keV).
Our final result for the integrated cross-section  requires to sum over all the shells with different binding energies $u_q$ :
\be
\label{eq:cscrystalFull}
\sigma = \sum_q \int_{k_{A,u_q}^{\mathrm{min}}}^{k_{A,u_q}^\mathrm{max}}dk_A\frac{\left|\mathcal{M}\right|^2k_A n_q(k_A) }{16\pi p_B|E_B k_A x_{0,u_q}(k_A)-E_{k_A} p_B|} ,
\ee
where $q = 1s, 2s, 2p- ,2p$\dots represent the DHF shells of  large $Z$ materials
described above, 
and $n_q (k) = (2\pi^2) \rho_q (k)$.

In practice, however, significant effects occur only  for the inner-most shells 
of large-$Z$ material
around the resonance (for which $k^\mathrm{min}_A$ tends to zero). While we included this effect for the $1s$ shell for tungsten, we found that 
in the not-yet excluded part of the parameter space the corresponding modification of the projected limits is negligible. As discussed in the main text this is simply the consequence of the fact that for masses around the resonance 
the signal mostly arises from weakly bound -- valence electrons -- orbitals, while for larger masses the correction terms in $k^\mathrm{min}_A$
become again negligible due to their suppression by $m_X^2$.

\section{Signal events in a thick target}
\label{app:signal-W}

The number of dark photon (DP) events in a thick target with atomic number $Z$ and mass number $A$ is given by~\cite{Nardi:2018cxi}:

\begin{equation}
N_{A'}=
\left(1-e^{\frac{z_D - z_{\mathrm{det}}}{\ell_\epsilon}}\right)
\frac{N_{\mathrm{poT}} N_\mathrm{Av}  X_0\rho}{A} 
\int_0^T dt\, \frac{d \mathcal{P}(t,z_D,\ell_\epsilon)}{dt}
\int dE_e \, \int dE \, \mathcal{G}(E, E_B, \sigma_{B}) \, I(E,E_e,t)\,\sigma (E_e),
\end{equation}
with  $N_{\mathrm{poT}}$ the number of positron on target, $N_\mathrm{Av}$ the Avogadro number, $X_0$ the radiation length ($X_0 = 3.5$ mm for tungsten), $T$  the target length in unit of radiation length and $\rho$ the density ($\rho = 19.3$ g/cm$^3$ for tungsten).
The first factor in parenthesis,
where $z_D$ is the target length, $z_\mathrm{det}$ the distance between the origin and the detector, and $\ell_\epsilon =  \gamma c \tau_{A'}$   the DP decay length,  subtracts out the $A'$s that decay after 
the detector and go undetected. Note that  
 the decay length depends explicitly on the atomic electron momentum via $\gamma=E_X(k)/m_X$,  
while $\tau_{A'}$ is the DP lifetime at rest given by 
$$
\tau_{A'}^{-1}=\Gamma_{A'}=\frac{\epsilon^2\alpha}{3} m_{A'} \left(1+2\frac{m_e^2}{m_{A'}^2}\right)\sqrt{1-4 \frac{m_e^2}{m_{A'}^2}}.
$$
The probability distribution 
 \begin{equation}
 \frac{ d \mathcal P(t,z_D,\ell_\epsilon)}{dt} =
 e^\frac{X_0 t-z_D}{\ell_\epsilon}\,,
\end{equation}
accounts for the number of DP that decay outside the target, 
given that most $e^+e^-$ produced inside the target get absorbed, while the Gaussian $\mathcal{G}(E, E_B, \sigma_B)$ describes the energy distribution of positron inside the beam. Finally, the probability of finding a positron with energy $E_e$ after passing through $t=z/X_0$ radiation lengths is~\cite{Bethe:1934za,Tsai:1966js}
\begin{equation}
    I(E,E_e,t)=\frac{\theta(E-E_e)}{E \Gamma(bt)}\left(\log\frac{E}{E_e}\right)^{bt-1},
\end{equation}
where $b=4/3$, $E$ is the initial positron energy  and $\Gamma$ is the Gamma function.

\bibliography{biblio} 
\end{document}